\documentclass[aps,prb,reprint,amsmath,amssymb,superscriptaddress,floatfix]{revtex4-2}
\usepackage{graphicx}
\usepackage[colorlinks, citecolor=blue, linkcolor=blue]{hyperref}
\usepackage{graphicx}
\usepackage{xcolor}
\usepackage{caption}
\usepackage{subcaption} 
\usepackage[normalem]{ulem}

\usepackage{epstopdf}
\graphicspath{
{/home/louise/Dropbox/images/matjes/nanoSkX/}
{/home/louise/Dropbox/images/matjes/J1J2/}
{/home/louise/Dropbox/images/matjes/nanoSkX/actual_skX/}
{/home/louise/Dropbox/images/matjes/nanoSkX/PdFeIr/}
{/home/louise/Dropbox/images/matjes/nanoSkX/PdFeIr/Jeff/}
{/home/louise/Dropbox/images/matjes/nanoSkX/PdFeIr/Jeff/gyration/}
{/home/louise/Dropbox/images/matjes/nanoSkX/PdFeIr/Jdft/}
{/home/louise/Dropbox/images/matjes/nanoSkX/PdFeIr/Jhoi/}
{/Users/louise/Dropbox/images/matjes/nanoSkX/}
{/Users/louise/Dropbox/images/matjes/J1J2/}
{/Users/louise/Dropbox/images/matjes/nanoSkX/actual_skX/}
{/Users/louise/Dropbox/images/matjes/nanoSkX/PdFeIr/}
{/Users/louise/Dropbox/images/matjes/nanoSkX/PdFeIr/Jeff/}
{/Users/louise/Dropbox/images/matjes/nanoSkX/PdFeIr/Jdft/}
{/Users/louise/Dropbox/images/matjes/nanoSkX/PdFeIr/Jhoi/}
}

\newcommand{\FNRS}{Fonds de la Recherche Scientifique (FRS - FNRS), B-1000 Brussels, Belgium}
\newcommand{\NANOMAT}{Nanomat/Q-mat/CESAM, Universit{\'e} de Li{\`e}ge, B-4000 Sart Tilman, Belgium}

\begin{document}

%
%

\title{Eigenmodes of magnetic skyrmion lattices}

\author{Louise Desplat}
\email{ldesplat@uliege.be}
\affiliation{\NANOMAT}

\author{Bertrand Dup{\'e}}
\affiliation{\NANOMAT}
\affiliation{\FNRS}

\date{\today}

%
%
\begin{abstract}
We explore the interplay between topology and eigenmodes by changing the stabilizing mechanism of skyrmion lattices (skX). We focus on two prototypical ultrathin films hosting an hexagonal (Pd/Fe/Ir(111)) and a square (Fe/Ir(111)) skyrmion lattice, which can both be described by an extended Heisenberg Hamiltonian.   We first examine whether the Dzyaloshinkskii-Moriya, or the exchange interaction as the leading energy term affects the modes  of the hexagonal skX of Pd/Fe/Ir(111). In all cases, we find that the lowest frequency modes correspond to  internal degrees of freedom of individual skyrmions, and suggest a classification based on azimuthal and radial numbers $(l,p)$, with up to $l=6$, and $p=2$. We also show that the gyration behavior induced by an in-plane field corresponds to the excitation of $l=1$ deformation modes with varying radial numbers. Second, we examine the square lattice of skyrmions of Fe/Ir(111). Its stabilization mechanism is dominated by the 4-spin interaction. After relaxation, the unit cell does not carry a topological charge, and the eigenmodes do not correspond to internal skyrmion deformations. By reducing the 4-spin interaction, the integer topological charge is recovered, but the charge carriers do not possess internal degrees of freedom, nor are they separated by energy barriers. We conclude that a 4-spin dominated Hamiltonian does not yield skyrmion lattice solutions, and that therefore,  \textit{a nontrivial topology does not imply  the existence of skyrmions}.
\end{abstract}

\maketitle

%
%

%
\section{Introduction}
Magnetic skyrmions are topologically nontrivial solitonic chiral spin textures localized in two dimensions at the nanometric scale~\cite{bogdanov1989thermodynamically,bogdanov1994thermodynamically}.  In systems with broken inversion symmetry, they are typically stabilized by the Dzyaloshinskii-Moriya interaction (DMI)~\cite{dzyaloshinskii,moriya} in competition with exchange, and anisotropies. 
Experimental observation of a skyrmion lattice (skX) phase in chiral magnets was first reported over a decade ago in bulk MnSi~\cite{muhlbauer2009skyrmion}. In skyrmion-hosting bulk magnets, the leading energy term responsible for spatially modulated spin configurations is the DMI, and the skX phase is stabilized at intermediate magnetic fields by the free energy, typically close to the critical temperature~\cite{bogdanov1994thermodynamically,butenko2010stabilization,yu2010real}.
Skyrmion lattices were later reported in ultrathin magnetic films \cite{romming2013writing}. In that case, density functional theory (DFT) calculations have shown that they are stabilized at zero temperature by the Gibbs energy, as a result of competing exchange, DMI, and anisotropy and/or magnetic field~\cite{dupe2014tailoring}.

Skyrmion lattices are especially attractive for applications in microwave electronics and nanomagnonics~\cite{chumak2015magnon}, whereby periodically arranged magnetic textures can be used to create magnonic crystals with reconfigurable wave properties~\cite{garst2017collective}. The nontrivial topology of the spin texture additionally results in the presence of topological magnon bands with nonzero Chern number, which can in turn create magnon edge-states, and be responsible for a magnon Hall effect~\cite{roldan2016topological,garst2017collective,weber2022topological}.
 As such, skyrmion lattices have been investigated in metallic (MnSi, FeGe), semiconducting (Fe$_{1-x}$Co$_x$Si,  GaV$_4$S$_8$), and insulating (Cu$_2$OSeO$_3$) chiral magnets~\cite{onose2012observation,okamura2013microwave,schwarze2015universal,kezsmarki2015neel,ehlers2016skyrmion}. These studies have pointed to a universal character of the skX eigenmodes, independently of the material~\cite{schwarze2015universal,garst2017collective}. In particular, in insulating materials, they  offer the potential for energy-efficient, high frequency wave-based computing technologies, with electric-field control of the  magnetic order and low spin-wave damping. For such applications, an in-depth understanding of the eigenmodes is necessary.
 
Besides the field of magnonics, studying the modes of skyrmionic systems gives insight into their fundamental properties such as thermal stability, or rigidity. The knowledge of eigenfrequencies is also useful for resonance experiments, e.g., to determine material parameters.

Localized modes of isolated skyrmions are typically found below the magnon continuum, and correspond to translation and $l$th order polynomial deformations of the skyrmion texture~\cite{makhfudz2012inertia,lin2014internal,schutte2014magnon,kravchuk2018spin}. These internal degrees of freedom were shown to be responsible for the skyrmion mass~\cite{makhfudz2012inertia}, and enhance its thermal stability through a large configurational entropy~\cite{desplat2018thermal,von2019skyrmion,desplat2020path}.

Meanwhile, in skyrmion lattices, three classes of  excitations were theoretically predicted~\cite{petrova2011spin,mochizuki2012spin,zhang2017eigenmodes} and  experimentally observed~\cite{onose2012observation,okamura2013microwave,schwarze2015universal,ehlers2016skyrmion}, namely, the (Goldstone) translation mode, clockwise (CW) and counterclockwise (CCW) gyration, and breathing. 
 Breathing is dynamically excited by an out-of-plane oscillatory magnetic field, while gyration is excited by an in-plane magnetic field. Gyration motion was shown to originate from the interplay of inertia and the emergent Lorentz force resulting from the topological magnetic texture~\cite{petrova2011spin}. The dispersion of the lowest energy magnon bands was derived theoretically, and some bands were shown to possess a nonzero Chern number~\cite{roldan2016topological,garst2017collective}, but the nature of these modes was not identified besides the three kinds mentioned above. Additionally, CW gyration is the only skyrmion mode which has been reported to possess a node in the radial direction~\cite{zhang2017eigenmodes,mruczkiewicz2017spin}.

In this article, we compute and classify the eigenmodes of magnetic skyrmion lattices. 
 We focus on transition metal thin films of Pd/Fe/Ir(111)~\cite{dupe2014tailoring} and Fe/Ir(111)~\cite{heinze2011spontaneous}, while the general results should hold for all chiral magnets.  The rest of this work is organized as follows.

In Sec.~\ref{sec:methods}, we first present the different formulations of the Heisenberg Hamiltonian used in this work, and we provide an overview of the method used to extract the  eigenmodes. 

Second, in Sec.~\ref{sec:disp}, we classify the sets of coefficients describing the magnetic properties of our ultra-thin films based on their different stabilization mechanisms. To do so, we compute the  energy dispersion of single spin spirals ($1Q$ states), and of the superposition of two spin spirals ($2Q$ states). We highlight the fact that, while in Pd/Fe/Ir(111), a minimum in the energy of single-$Q$ spirals is created close to the $\overline{\mathrm{\Gamma}}$ point $(q=0)$ of the first Brillouin zone (BZ) by the interplay of exchange and DMI~\cite{dupe2014tailoring}, in Fe/Ir(111), the competition of exchange and the 4-spin interaction creates an energy minimum for 90-degree spin spirals around the middle of the BZ~\cite{heinze2011spontaneous}.

Third, the lowest frequency modes of the skX ground state of Pd/Fe/Ir(111) are derived In Sec.~\ref{sec:zoology}. We suggest a classification of the modes based on $(l,p)$ azimuthal and radial numbers. We find that the nature of the low frequency skX modes as internal skyrmion deformations is independent of the formulation of the Hamiltonian.
Next, in Sec.~\ref{sec:feir}, we examine the modes of the ground state of Fe/Ir(111), the so-called nanoskyrmion lattice,  as well as that of a fictitious system obtained by reducing the 4-spin amplitude by half.  We find that the 4-spin interaction can stabilize a lattice of topological objects which are not skyrmions, as they do not possess internal degrees of freedom, and are not separated by energy barriers. This demonstrates that a topological charge does not guarrantee the existence of skyrmions, and that neither energy barriers nor internal degrees of freedom automatically derive from the topology.

After that, in Sec.~\ref{sec:dyn}, we perform magnetization dynamics simulations and show that selective modes can be excited depending on the azimuthal number carried by an applied magnetic field. We identify the CCW and CW modes as $l=1$ deformation modes with amplitude localized respectively far from, and onto the skyrmion core.

Last, the results are summarized in Sec.~\ref{sec:cl}, and some perspectives are discussed.

%
%
\section{Model and methods}\label{sec:methods}
%
\paragraph{Magnetic Hamiltonian}
We simulate $N$ magnetic moments $\mathbf{M}=\{\hat{\mathbf{m}}_i\}$ of norm unity on a hexagonal lattice with periodic boundary conditions. Atomistic simulations are performed with the Matjes code~\cite{matjes}, and the Spirit atomistic framework~\cite{muller2019spirit}. The Heisenberg Hamiltonian used throughout this work has the general form:
  \begin{equation}\label{eq:hamil}
 \mathcal{H}  =  \mathcal{H}_{\mathrm{ex}} - \sum_{ij} \mathbf{D}_{ij} \cdot
 \left( 
 	\mathbf{\hat{m}}_i \times \mathbf{\hat{m}}_j
 \right) 
  - K \sum_i m_{z,i}^2  - \mu_s \sum_i \mathbf{B} \cdot \mathbf{\hat{m}}_i, 
\end{equation}
where $\mathcal{H}_{\mathrm{ex}}$ contains contribution from the Heisenberg exchange and higher-order terms,  $\mathbf{D}_{ij}$ is the interfacial DMI vector between first neighbors $i$ and $j$, $K$ is the effective perpendicular magnetic anisotropy constant, and $\mathbf{B}$ is the external applied magnetic field. We neglect demagnetizing fields, as it was shown that the effect of the dipole-dipole interaction on the energy landscape in ultrathin films can be well approximated by an effective anisotropy~\cite{lobanov2016mechanism}.

For $\mathcal{H}_{\mathrm{ex}}$, we use three different formulations: 
\begin{itemize}
\item Effective Heisenberg exchange:
  \begin{equation}\label{eq:J_eff}
\mathcal{H}_{\mathrm{ex}}^{\mathrm{eff}}= - J_{\mathrm{eff}} \sum_{ij} 
 \left(
 	 \mathbf{\hat{m}}_i  \cdot \mathbf{\hat{m}}_j
 \right),
\end{equation}
in which $J_{\mathrm{eff}}$ is the effective isotropic exchange coupling between first nearest neighbors; 

\item Extended Heisenberg exchange~\cite{dupe2014tailoring}:
  \begin{equation}\label{eq:J_dft}
\mathcal{H}_{\mathrm{ex}}^{\mathrm{ext}}= -  \sum_{ij} J_{ij}
 \left(
 	 \mathbf{\hat{m}}_i  \cdot \mathbf{\hat{m}}_j
 \right),
\end{equation}
in which $J_{ij}$ extends beyond the first nearest neighbours;

\item Extended Heisenberg exchange and high-order interactions (HOI)~\cite{heinze2011spontaneous}:
  \begin{equation}\label{eq:J_hoi}
\begin{split}
& \mathcal{H}_{\mathrm{ex}}^{\mathrm{HOI}}  =   -  \sum_{ij} J_{ij} 
 \left(
 	 \mathbf{\hat{m}}_i  \cdot \mathbf{\hat{m}}_j
 \right)
  - \sum_{ij} \mathcal{B}_{ij}
 \left(
 	\mathbf{\hat{m}}_i \cdot \mathbf{\hat{m}}_j
 \right)^2 \\
 & - \sum_{ijkl} \mathcal{K}_{ijkl} 
 \big[
 	\left(
 		\mathbf{\hat{m}}_i \cdot \mathbf{\hat{m}}_j
 	\right)
 	\left(
 		\mathbf{\hat{m}}_k \cdot \mathbf{\hat{m}}_l
 	\right)	
 	+
 	\left(
 		\mathbf{\hat{m}}_i \cdot \mathbf{\hat{m}}_l
 	\right)
 	\left(
 		\mathbf{\hat{m}}_j \cdot \mathbf{\hat{m}}_k
 	\right) \\	
 	 & - 
 	 \left(
 		\mathbf{\hat{m}}_i \cdot \mathbf{\hat{m}}_k
 	\right)
 	\left(
 		\mathbf{\hat{m}}_j \cdot \mathbf{\hat{m}}_l
 	\right)	
 \big], \\
 \end{split}
\end{equation}
where $\mathcal{B}_{ij}$, and $\mathcal{K}_{ijkl}$  are respectively the biquadratic, and four-spin interaction exchange constants. Here, the biquadratic interaction is limited to first nearest neighbors, and the 4-spin interaction, to the first nearest quadruplets. 
\end{itemize}

\paragraph{Extracting eigenmodes} The eigenmodes of the dynamics are derived in the harmonic approximation. The Hamiltonian in Eq.~(\ref{eq:hamil}) is linearized by expanding it in second order of small deviations from the ground state. The result is then injected into the dynamics equations.  We obtain a set of $N$ eigenfrequencies $\{\omega_k\}$ and corresponding eigenvectors $\{\boldsymbol{\chi}_k\}$, where $k=1 \hdots N$ is the mode index.  More details are given in Appendix~\ref{app:extract_modes}.


\section{Stabilization mechanism}\label{sec:disp}
\begin{figure}
\includegraphics[width=1\linewidth]{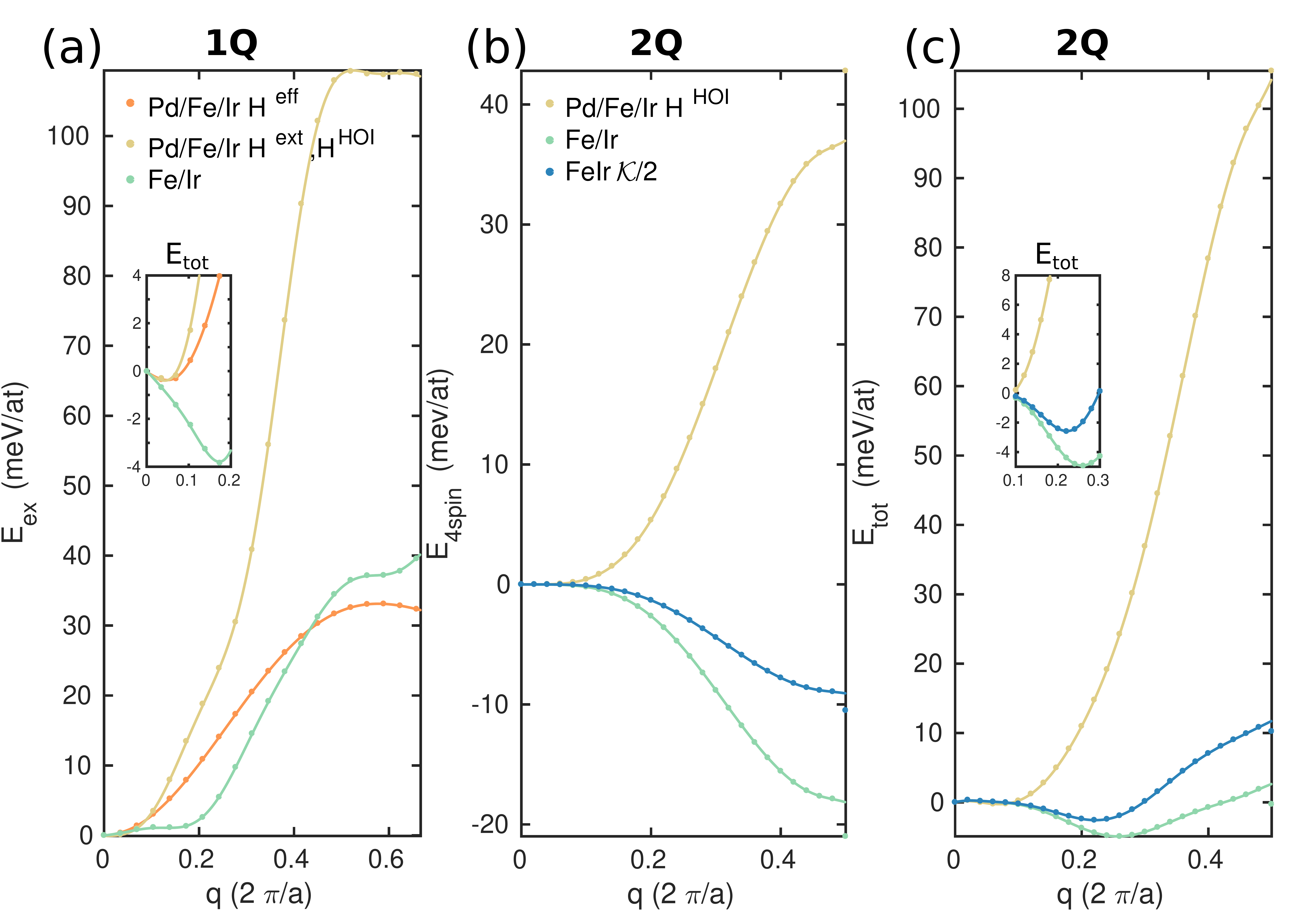}
\caption{Energy dispersions of spin spirals for Pd/Fe/Ir(111), Fe/Ir(111), and Fe/Ir(111) with reduced 4-spin interaction. (a) Exchange energy dispersion of $1Q$ spin spirals along the high-symmetry line $\overline{\mathrm{\Gamma K}}$. The inset shows a closeup of the total energy close to $\overline{\mathrm{\Gamma}}$. (b, c) Energy dispersion of (b) the 4-spin interaction and (c) the total energy of 90-degree $2Q$ spin spirals propagating along $\overline{\mathrm{\Gamma M}}$ and $\overline{\mathrm{\Gamma K}}$. The inset in (c) shows a closeup of the total energy around the center of the Brillouin zone. The zero of the energy is chosen as that of the ferromagnetic state $(q=0)$. The lines are spline intended as a guide to the eye.} 
\label{fig:dispersions}
\end{figure}

 \paragraph{$1Q$ dispersions}
 In Figs.~\ref{fig:dispersions}a, we show the energy dispersion of $1Q$ N\'eel spin spirals propagating along the $\overline{\Gamma\mathrm{K}}$ direction, in Pd/Fe/Ir(111) at zero magnetic field with the three formulations of the Hamiltonian (Eqs. (\ref{eq:hamil})-(\ref{eq:J_hoi}))~\cite{dupe2014tailoring,von2017enhanced,paul2020role}, and in Fe/Ir(111)~\cite{heinze2011spontaneous}.  More details are given in Appendix~\ref{app:dispersions}. The $1Q$ dispersion only  depends on the exchange, the DMI, the anisotropy, and the biquadratic energies. In this case, the 4-spin interaction does not play any role, as its energy contribution is $-12 \mathcal{K}$ for all single-$Q$ states. In what follows, wave vectors are expressed in units of $2\pi/a $, where $a=2.7 \AA$ is the lattice constant of Fe. Several cases have to be distinguished. 

First, in Pd/Fe/Ir(111) with $\mathcal{H}_{\mathrm{ex}}^{\mathrm{eff}}$, the exchange is almost quadratic close to $\overline{\mathrm{\Gamma}}$. When, instead, the extended exchange term $\mathcal{H}_{\mathrm{ex}}^{\mathrm{ext}}$ is used to describe the system, the exchange energy leads to a very flat dispersion up to $q \sim 0.05$,  which implies that a long-range noncollinear state such as a spin spiral costs very little exchange energy. Depending on the fitting parameters, the energy of the extended Heisenberg model can even exhibit a small energy minimum~\cite{dupe2014tailoring}. This is in stark contrast to the effective Hamiltonian model. The difference of behavior close to the $\overline{\mathrm{\Gamma}}$-point explains the large discrepancy in the energy at the edge of the BZ. The DMI splits the energies of left- and right-rotating spin spirals, and yields a minimum in the total energy around $q \sim 0.05$ for right-rotating spin spirals. Note that the contribution of the biquadratic term is equivalent to a change in the 3rd-neighbor exchange coupling $J_3$, so the sum of the exchange and the biquadratic contributions in $\mathcal{H}_{\mathrm{ex}}^{\mathrm{HOI}}$ yields the same energy as the exchange in $\mathcal{H}_{\mathrm{ex}}^{\mathrm{ext}}$.

In the case of Fe/Ir(111), the dispersion is flat up to $q\sim 0.2$, which leads to zero effective exchange. In that case, the DMI plays a major role, as it favors a 90-degree angle between neighboring magnetic moments, corresponding to a minimum at $q=0.25$. When the DMI is taken into account, the minimum in the energy of the spin spirals is found at $q \sim 0.17$.

 \paragraph{$2Q$ dispersions}
When higher-order magnetic interactions, such as the 4-spin interaction, are taken into account, the exploration of the stabilization mechanisms becomes more complex. The 4-spin interaction has a constant dispersion for 1$Q$ spin spiral states. It is then necessary to explore a 2$Q$ spin spiral dispersion, i.e., the superposition of two spin spirals, as described in Heinze \emph{et al.}~\cite{heinze2011spontaneous}. Since the 4-spin interactions is minimized for a 90-degree angle between wave vectors~\cite{heinze2011spontaneous}, we restrict the dispersion to the spin spirals propagating in the $\mathbf{q}_1 \parallel \overline{\mathrm{\Gamma M}}$ and $\mathbf{q}_2 \parallel \overline{\mathrm{\Gamma K}}$ directions with $\mathbf{q}_1 \perp \mathbf{q}_2$ for $q_{1,2} \in [0,0.5]$, in units of $2\pi/a_{1,2}$. Further details are given in Appendix~\ref{app:dispersions}.

For Pd/Fe/Ir(111), the dispersion of 2$Q$ spin spirals is qualitatively similar to that of 1$Q$ for exchange and DMI, but the 4-spin interaction increases with $q$ and reaches a maximum at the edge of the BZ ($q_1 = q_2 = 0.5$). On the other hand, in Fe/Ir(111), the 4-spin interaction has the opposite sign, and is minimum at the edge of the BZ. In the end, the contribution of the 4-spin creates a lower minimum for 2Q states in Fe/Ir around $q_1 = q_2 \sim 0.26$, with $E_{\mathrm{tot}}=-4.92$~meV/at. below the FM state. When the 4-spin strength is reduced by half, this minimum is moved  towards $\mathrm{\overline{\Gamma}}$, at $q\sim 0.22$.

In summary, while the interplay of exchange and DMI creates a minimum for single-$Q$ spin spirals close to $\overline{\mathrm{\Gamma}}$ in Pd/Fe/Ir(111), in Fe/Ir(111) it is the interplay of exchange and the 4-spin interaction which creates a lower minimum for a combination of 90-degree spin spirals around the middle of the BZ. The implications of this observation for noncollinear magnetic states in these systems will be uncovered in the rest of this work.

\section{Eigenmodes of a skX stabilized by exchange and DMI}\label{sec:zoology}

In this section, we focus on Pd/Fe on Ir(111), a skyrmion-hosting system that has been extensively studied  both theoretically~\cite{dupe2014tailoring,von2017enhanced,bottcher2018b} and experimentally~\cite{romming2013writing,romming2015field}. 
At zero temperature, the system exhibits a skyrmion lattice ground state at intermediate magnetic fields, which persists until around $80$~K~\cite{bottcher2018b,lindner2020temperature}. Later on, it was shown that energy barriers of isolated skyrmions in this system were sensitive to the inclusion of the 4-spin interaction in the Hamiltonian~\cite{paul2020role}.

In what follows, we investigate the lowest frequency eigenmodes of the skyrmion lattice of Pd/Fe/Ir(111) under three different formulations of the Hamiltonian from Eqs.~(\ref{eq:hamil})-(\ref{eq:J_hoi}), namely, effective exchange, extended exchange, and extended exchange with higher-order terms. In particular, the 4-spin interaction in the later has a value of $\mathcal{K}=2.14$~meV/at~\cite{paul2020role}. Each Fe atom carries a magnetic moment $\mu_S=3 \mu_B$, where $\mu_B$ is the Bohr magneton. The damping is set to $\alpha=0.3$. The supercell contains $N=60\times60$ atomic sites for $\mathcal{H}_{\mathrm{ex}}^{\mathrm{eff}}$ and $\mathcal{H}_{\mathrm{ex}}^{\mathrm{ext}}$, and  $N=65\times65$ for $\mathcal{H}_{\mathrm{ex}}^{\mathrm{HOI}}$.

\subsection{The skyrmion lattice ground state}
\begin{figure}
\includegraphics[width=.7\linewidth]{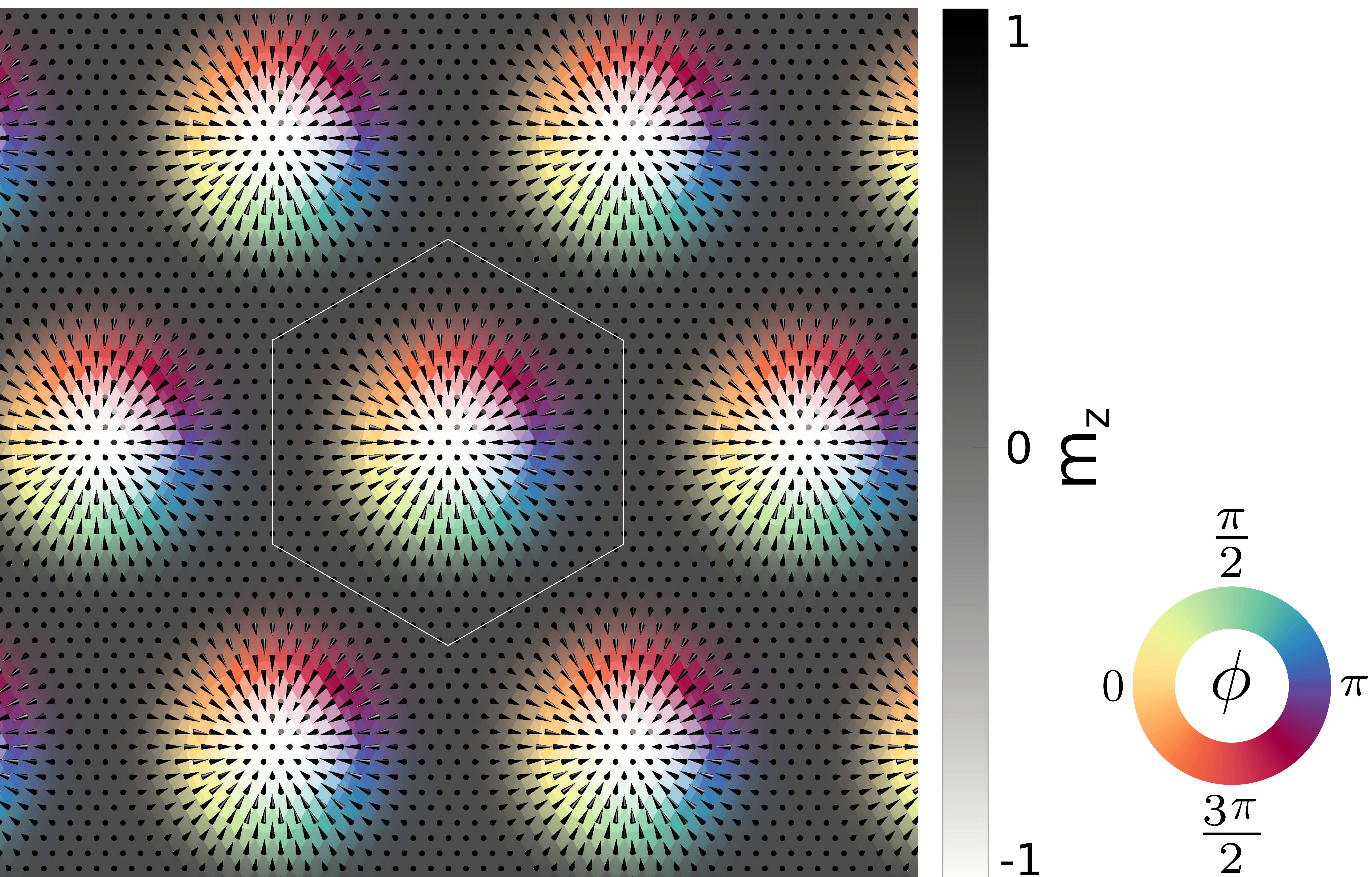}
\caption{Relaxed portion of the skyrmion lattice ground state for $\mathcal{H}_{\mathrm{ex}}=\mathcal{H}_{\mathrm{ex}}^{\mathrm{eff}}$ and $B_z=2.5$~T. The unit cell is shown in white. 
} 
\label{fig:skX}
\end{figure}

First, the skX ground state is relaxed with overdamped spin dynamics simulations~\cite{garcia1998langevin} for all three formulations of the Hamiltonian. We set the out-of-plane magnetic field to $B_z=2.5$~T, corresponding to the skX phase for all three Hamiltonians~\cite{von2017enhanced,paul2020role}.
The relaxed skX  are very similar, with wave vector $q_{\mathrm{sk}} = 0.05$ for $\mathcal{H}_{\mathrm{ex}}^{\mathrm{eff}}$ and $\mathcal{H}_{\mathrm{ex}}^{\mathrm{ext}}$, and $q_{\mathrm{sk}} = 0.04$ for $\mathcal{H}_{\mathrm{ex}}^{\mathrm{HOI}}$. 
The larger wavelength with HOI is coherent with the fact that the higher-order terms were shown to increase the radii of isolated skyrmions in this system~\cite{paul2020role}.
 In Fig. \ref{fig:skX}, we show a portion of the relaxed skX for $\mathcal{H}_{\mathrm{ex}}^{\mathrm{eff}}$, where the unit cell is shown in white. 
%

\subsection{$(l,p)$ mode classification}
Next, the modes are extracted as described in App.~\ref{app:extract_modes}. Their profiles are characterized by the real part of the polar $\theta$ components of the eigenvectors $\boldsymbol{\chi}_k$, which amounts to setting the out-of-plane $z$ direction as the quantization axis. We suggest a classification of the uniform modes according to their $(l,p)$ numbers. $l$ is the azimuthal number, such that $2l$ nodes are encountered when travelling around a skyrmion in the azimuthal direction. $p$ is the radial number and gives the number of nodes the radial direction. 

\begin{figure*}
\includegraphics[width=1\linewidth]{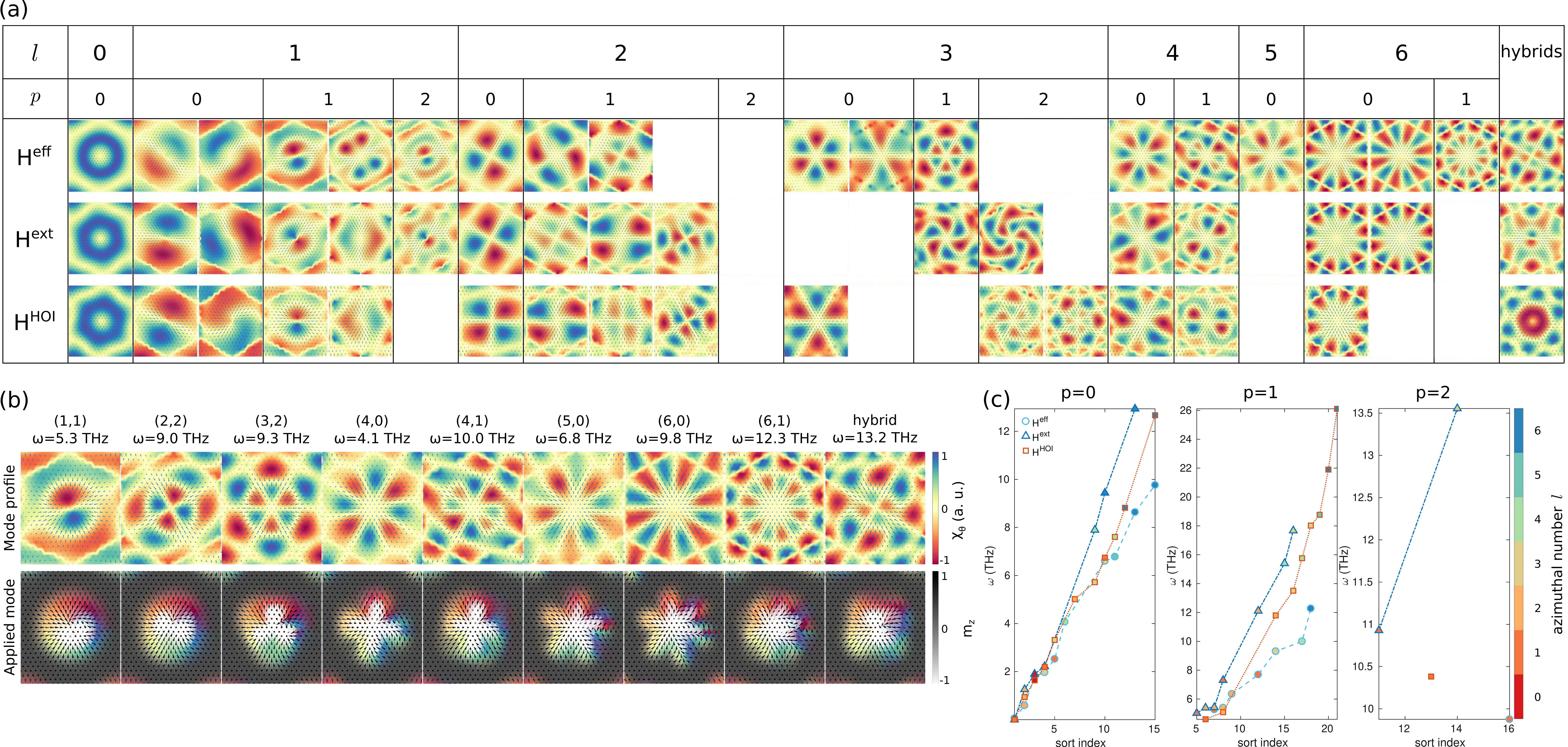}
     \hfill  
\caption{Lowest frequency uniform modes of the skyrmion lattice ground state of Pd/Fe/Ir(111) in the harmonic approximation, classified by azimuthal and polar numbers $(l,p)$. (a) $\theta$ profiles of the eigenvectors of the first 200 lowest frequency uniform modes, classified by increasing $l$ and $p$. The three rows correspond to the three different formulations of the Hamiltonian given in Eqs.~(\ref{eq:hamil})-(\ref{eq:J_hoi}) . (b) Example of modes for effective Heisenberg exchange, where the top row shows the $\theta$ profile of the eigenvector, and the bottom row shows the magnetic texture resulting from the application of the mode to the skX according to Eq.~(\ref{eq:apply_mode}). The  colorcode is the same  as that of Fig.~\ref{fig:skX}. The amplitudes are set to $A_0=50$ or 100 for better visibility. In all mode profiles, the ground state is superimposed as a guide to the eye. The view is limited to one unit cell. (c) Eigenfrequencies of the uniform modes shown in (a) sorted by increasing value, where each subplot corresponds to a different $p$ branch. The different formulations of $\mathcal{H}_{\mathrm{eff}}$ are indicated by the color of the lines and marker shape, and the azimuthal number $l$ is given by the color inside the markers.}
 \label{fig:PdFeIr_modes} 
\end{figure*}

The results are gathered in Fig.~\ref{fig:PdFeIr_modes}.
Similarly to an isolated skyrmion state, the lowest frequency modes correspond to coupled internal deformations of the individual skyrmions, and are either uniform, i.e., all the skyrmions are deformed in the same way, or nonuniform. In nonuniform modes, either different types of internal modes are excited, such as, for instance, translation and elliptic deformation, or the same mode is excited along different axes for different skyrmions. In the following, we focus on uniform modes amongst the first 200 lowest frequencies, which, in Pd/Fe/Ir(111), corresponds to the $10^8- 10^{13}$~Hz range.

Fig.~\ref{fig:PdFeIr_modes}a shows the $\theta$ profiles of the uniform modes for all three formulations of the Hamiltonian, ordered by increasing $l$ and $p$ numbers. The corresponding frequencies are given in Fig.~\ref{fig:PdFeIr_modes}c for each $p$ branch, where the azimuthal number $l$ is indicated by the color inside the markers. 
In all cases, the lowest frequency modes are the skyrmion deformation modes that are commonly reported in isolated skyrmions: two translation modes with $(l,p)=(1,0)$, breathing $(0,0)$, as well as elliptical $(2,0)$, and triangular $(3,0)$ deformations. Note that the low-frequency translation mode is not gapless, but possesses a finite frequency in the 100~MHz-10~GHz range due to the weak pinning of the skyrmions to the crystal lattice. The faster $(1,0)$ mode is found around 2~THz. The presence of both low- and high-frequency $(1,0)$ modes in the skX is in agreement with theorectical predictions~\cite{petrova2011spin}.

Examples of modes beyond these more common ones are shown in Fig.~\ref{fig:PdFeIr_modes}b, for $\mathcal{H}_{\mathrm{ex}}=\mathcal{H}_{\mathrm{ex}}^{\mathrm{eff}}$. The top row corresponds to the $\theta$ profiles of the eigenvectors, while the bottom row shows the spin configuration that results from the application of the mode to the skX ground state, according to Eq.~(\ref{eq:apply_mode}).
We find, on the one hand, higher-order azimuthal deformations: square $(4,0)$, pentagonal $(5,0)$, and hexagonal $(6,0)$ modes. Such modes were previously reported for isolated skyrmions in the skX phase at low magnetic field or perpendicular anisotropy, while typically not being physically accessible due to the elliptic instability~\cite{lin2014internal}.
On the other hand, we report higher-order radial modes up to $p=2$. When these modes are excited, the core $\{\mathbf{m}_i |  m_{z,i}\lesssim 0\}$, the envelope $\{\mathbf{m}_i | m_{z,i} \approx 0\}$, and the tail $\{\mathbf{m}_i |  m_{z,i}
\gtrsim 0  \}$ of the skyrmions can be deformed in different ways. For instance, the $(1,1)$ mode, shown in the first column in Fig.~\ref{fig:PdFeIr_modes}b, results in  the antiphase translation of the core and tail of the skyrmions in opposite directions.

Additionally, we find hybrid modes, examples of which are shown in the last columns of Figs.~\ref{fig:PdFeIr_modes}a and b. In this case, the azimuthal number varies in the radial direction. For instance, the hybrid mode in Fig.~\ref{fig:PdFeIr_modes}b, has $p=1$, with $l_{p=0}=2$ and $l_{p=1}=4$. When this mode is excited, the core undergoes elliptical deformation, while the envelope and tail undergo square deformation.

This classification highlights an interesting ressemblance of the skyrmion modes with atomic orbitals, where, for a given $l$,  the energy (frequency) increases with radial number $p$.  A third quantum number, $m$--the magnetic number, could be used to differentiate between modes with the same $(l,p)$ values and different orientations. 

Nevertheless, the profiles of the higher frequency modes [Fig.~\ref{fig:PdFeIr_modes}a] 
hint at the fact that this classification is more valid at low frequency,  where the number of nodes remains low.
With a higher number of nodes, the hexagonal symmetry of the system is more prevalent, and  mode profiles often no longer resemble  solutions with cylindrical symmetry. For instance, the last $(2,1)$ modes shown for $\mathcal{H}_{\mathrm{ex}}^{\mathrm{ext}}$ and $\mathcal{H}_{\mathrm{ex}}^{\mathrm{HOI}}$ only possess  antinodes with $p=1$ along a single axis. Additionally,  most of the $l=4$ modes exhibit only a 2-fold symmetry and appear to be a superposition of $l=2$ modes along orthogonal axes. Such discrepancies are more pronounced for those classes of modes, as 2-fold and 4-fold symmetries are harder to accomodate onto the underlying hexagonal symmetry of the system, compared to the 2-, 3- and 6-fold ones. For the same reason, modes with $l=5$ are almost nonexistent.

\subsection{Effect of frustrated exchange and HOI}
The influence of exchange frustration and higher-order terms on the eigenfrequencies is visible on the $p=0$ and 1 branches in Fig.~\ref{fig:PdFeIr_modes}c. At low  frequency (low $l$), the graphs are almost superimposed, and so the formulation of $\mathcal{H}_\mathrm{ext}$ has a negligible effect on these modes. With increasing frequency (increasing $l$), $\mathcal{H}_{\mathrm{ex}}^{\mathrm{ext}}$ yields higher frequencies than $\mathcal{H}_{\mathrm{ex}}^{\mathrm{eff}}$. This is coherent with the fact that, with the inclusion of frustrated exchange, the  coupling to the crystal lattice increases, and so do the energy scales. Interestingly, the inclusion of HOI yields lower frequencies than $\mathcal{H}_{\mathrm{ex}}^{\mathrm{ext}}$, which may be due to the larger skyrmion size with HOI. In general, the effect of extended exchange and HOI on the higher frequencies seems more pronounced as $p$ increases. 

The effect of the formulation of $\mathcal{H}_{\mathrm{ex}}$ is also visible in the mode profiles, whereby the enhanced coupling to the crystal lattice with frustrated exchange and HOI results in more dramatic breaking of the  cylindrical symmetry of the mode profiles. For instance, the $(0,0)$ breathing mode acquires a more hexagonal profile for   $\mathcal{H}_{\mathrm{ex}}^{\mathrm{ext}}$ and $\mathcal{H}_{\mathrm{ex}}^{\mathrm{HOI}}$, and the symmetry of the $l=2$ and $l=4$ modes are also more reduced than for  $\mathcal{H}_{\mathrm{ex}}^{\mathrm{eff}}$.

In summary,  in Pd/Fe/Ir(111), where noncollinear states are stabilized by the interplay of exchange and DMI, the low frequency modes of the skX state correspond to coupled $(l,p)$ deformations of the individual skyrmions, reminiscent of atomic orbitals. This is independent of the inclusion of frustrated exchange and higher-order terms. At higher frequencies, modes with more nodes tend to lose their  cylindrical symmetry, and the $(l,p)$ classification appears less pertinent. The inclusion of exchange frustration and HOI does not affect the lower frequencies, while it tends to increase the larger ones--and moreso for a larger $p$ number. 

In Sec.~\ref{sec:dyn}, we will show how these modes can be dynamically excited with a magnetic field matching their azimuthal number, and we will identify and explain the observation of (C)CW modes.  Before that,  in the next section, we examine the modes of skX states stabilized by the interplay of exchange and the 4-spin interaction.

%
\section{ Eigenmodes of a skX stabilized by exchange and the 4-spin interaction}\label{sec:feir}

\subsection{The ground state of Fe/Ir(111)}

In Fe/Ir(111), the strong 4-spin interaction favors multi-$Q$ modulated states over single-$Q$ states. In combination with the DMI that selects a particular sense of rotation of the magnetization, its interplay with exchange leads to  the peculiar nanoskyrmion lattice (nanoskX) ground state of this system~\cite{heinze2011spontaneous}.

Following  parameters derived from first-principles~\cite{heinze2011spontaneous}, the Hamiltonian has the form in Eqs.~(\ref{eq:hamil}) and (\ref{eq:J_hoi}), with a 4-spin interaction amplitude of $\mathcal{K}=-1.05$~meV, and zero applied magnetic field. We simulate a single unit cell of $N=15 \times 15$. Each Fe atom caries a magnetic moment of amplitude $\mu_s=2.7\mu_B$. Note that the reduction of magnetic polarization compared to Pd/Fe/Ir(111) is due to Pd, which brings an extra contribution of $0.3\mu_B$ in the latter~\cite{von2017enhanced}.

\begin{figure}
 	\includegraphics[width=1\linewidth]{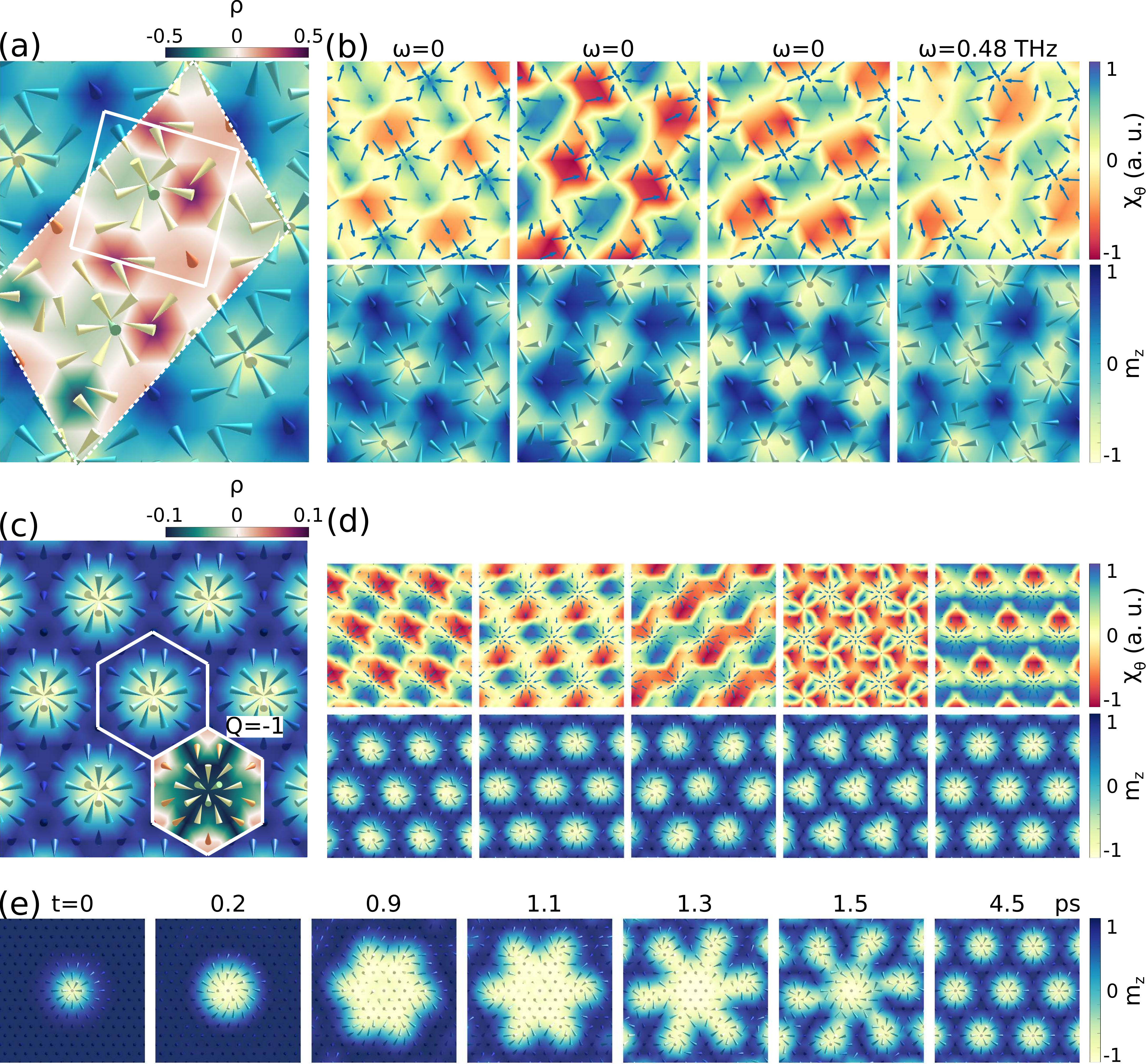}	
\caption{(a, c) Relaxed portion of the skyrmion lattice ground state of (a) Fe/Ir(111), (c) Fe/Ir(111) with the 4-spin interaction reduced by half. The  pseudo magnetic unit cell is sketched in white solid lines. The insets show the topological charge density over one pseudo unit cell.  In (a), the pseudo unit cell of the topological charge density is also indicated in white dashed lines. (b, d) Examples of low-frequency uniform modes with $\alpha=1$, where the top row shows the mode $\theta$ profile, and the bottom row shows the same mode applied to the ground state according to Eq.~(\ref{eq:apply_mode}). In the top row, the relaxed magnetic texture is superimposed as a guide to the eye. The scaling factor is set to (b) $A_0=5$, (d) $A_0=25$. (e) Snapshots of the dynamics of the nucleation of a topologically nontrivial multi-$Q$ state from a single initial skyrmion with $\alpha=0.5$ and $\mathcal{K}=-0.53$~meV/at.}
\label{fig:nanoskX}
\end{figure}

A portion of the relaxed nanoskX is shown in Fig.~\ref{fig:nanoskX}a. It has an energy of -9.97~meV/at. with respect to the ferromagnetic state, which is coherent with the value of -7~meV/at. given in Ref. ~\onlinecite{heinze2011spontaneous} for the unrelaxed state. Note that the real magnetic unit cell is the entire simulated supercell, while the pseudo magnetic unit cell is sketched in white solid lines in Fig.~\ref{fig:nanoskX}a.

To characterize the structure, we compute its discrete topological charge $Q$~\cite{berg1981definition,bottcher2018b}. The topological charge density $\rho$ is shown in Fig.~\ref{fig:nanoskX}a, in the pseudo unit cell delimited by the dashed line. This is because the topological charge density does not exhibit the quasi square periodicity of the magnetic texture.  We find that the square unit cell carries of a topological charge of $Q=0.2$, and is therefore not a lattice of skyrmions, but rather, based on the topological charge distribution, a lattice of bimerons with alternating polarisation.

Next, some of the pseudo-uniform mode profiles and the result of their application to the ground state are shown in Fig.~\ref{fig:nanoskX}b.  The damping of the system at cryogenic temperatures at which the nanoskX remains stable was obtained by first principles calculation around $\alpha=0.3$~\cite{yang2022radio}. We set $\alpha=1$ for sharper looking mode profiles. In this overdamped regime, the system possesses some zero-frequency modes, as shown in Fig.~\ref{fig:nanoskX}b. When excited, they simply decay exponentially in time, following Eq.~(\ref{eq:apply_mode}) with $\omega_k=0$. We note that these are however not Goldstone modes, because the system has no flat energy curvature. In the underdamped regime, they recover an oscillatory behavior in the THz range.
Unlike in Pd/Fe/Ir(111), we find that the lower frequency mode amplitudes are not consistently localized onto the "skyrmions", and do not correspond to internal deformation modes. An exception is a mode akin to $(1,0)$ shown in the first column of Fig.~\ref{fig:nanoskX}b. It  leads to the translation of the whole texture along the crystal lattice unit vector $\mathbf{a}_1=(1/2, -\sqrt{3}/2)$ that coincides with the  diagonal of the magnetic lattice. Nevertheless, when this mode, and all the others, are applied to the ground state, the skyrmion-like texture is destroyed, and the fractioned topological charge is not conserved.

 Based on these arguments, we conclude that the ground state of Fe/Ir(111) as obtained after relaxation of our supercell, is, in fact, not a skyrmion lattice, but rather a multi-$Q$ state driven by the 4-spin interaction. However, since the magnetic pseudo unit cell  does not carry a topological charge, this result remains in agreement with the conclusions of Sec.~\ref{sec:zoology}, i.e., the ground state does not contain skyrmions, and so its eigenmodes do not correspond to internal skyrmion deformations.
 
\subsection{Reduced 4-spin interaction}

 In the following, we reduce the 4-spin interaction by half, $\mathcal{K'}=-0.53$~meV/at., while keeping all the other parameters the same, and the size of the supercell is increased to $N=30 \times 30$. 
 
 The relaxed state is shown in Fig.~\ref{fig:nanoskX}c, and has the form of a hexagonal skyrmion lattice, with $Q=-1$ per unit cell., and wave vector $q_{\mathrm{sk}} = 0.2 $. It has a total energy of $-6.10$~meV/at. with respect to the FM state, and is indeed lower than the minima in both $1Q$ and $2Q$ dispersions, which respectively correspond to total energies of -3.8~meV/at., and -2.6~meV/at.  [Figs.~\ref{fig:dispersions}a, c]. Note that in this case, the minimum of the $2Q$ states is above that of the $1Q$ states, but a lower minimum should be found for a superposition of 3 spin spirals with equilateral wave vectors yielding a hexagonal skyrmion lattice ~\cite{muhlbauer2009skyrmion}.
 
We once more derive the eigemmodes of this new state. Examples of lower-frequency uniform modes are shown in Fig.~\ref{fig:nanoskX}d. Surprisingly, we do not recover skyrmion deformation modes. The amplitude of the modes is not localized onto the topological charge carriers, and they do not correspond to internal deformation of the $(l,p)$ nature, besides the first mode akin to $(1,0)$ translation in  Fig.~\ref{fig:nanoskX}d. 

Next, in Fig.~\ref{fig:nanoskX}e, we show snapshots of the dynamics of the system over 5~ps when initialized with a single isolated skyrmion, and $\alpha=0.5$. Surprisingly, the topologically nontrivial lattice nucleates spontaneously from the single skyrmion, i.e., topological charge creation occurs without having to overcome energy barriers.

Therefore, despite the non-trivial topological charge, this state appears to only be a multi-$Q$ state, but not a skyrmion lattice. The charge carriers do not behave as individual entities, as i. they do not possess internal degrees of freedom, and ii. they are not separated by energy barriers. This shows that a topological charge is not enough to ensure that a magnetic texture is a skyrmion, and that the energy barrier separating a skyrmion from other states does not automatically derive from the topology.

In the present system, even though we reduced the 4-spin interaction, it still favours multi-$Q$ over single-$Q$ states. When it is reduced further, the single-$Q$ spin-spiral ground state created by the interplay of Heisenberg exchange and DMI is recovered [Fig.~\ref{fig:dispersions}a], and either an out-of-plane magnetic field, or an increased perpendicular anisotropy, is required to yield a skX state. In this state, the skyrmions are very small but do possess the $(0,0)$ and $(1,0)$ modes, and they are separated  by energy barriers.

In summary, in systems where noncollinear magnetic states are stabilized by exchange and the 4-spin interaction rather than the DMI, magnetic textures with a nontrivial topology may not be skyrmions. We have found that the 4-spin interaction can stabilize lattices of topological objets which are not skyrmions.

%
\section{Dynamical SkX mode excitation}\label{sec:dyn}

In this section, we perform magnetization dynamics simulations on the skX state of Pd/Fe/Ir(111) [Fig.~\ref{fig:skX}], with a time-varying magnetic field, by numerical integration of Eq.~(\ref{eq:LLG})~\cite{garcia1998langevin}. We use the  Hamiltonian in Eq.~(\ref{eq:hamil}) with $\mathcal{H}_{\mathrm{ex}}^{\mathrm{eff}}$, while the upcoming results should hold for all three formulations of  $\mathcal{H}_{\mathrm{ex}}$. The damping is set to a more realistic value of $\alpha=0.01$. 
In the following, we demonstrate selective mode excitation based on their azimuthal number $l$. The modes obtained in the dynamics are in good agreement with the ones obtained in the harmonic approximation~[Eq.~(\ref{eq:linear_llg})].
Additionally, we reproduce the CCW and CW gyration behavior initially described in Ref.~\onlinecite{mochizuki2012spin} and explain its origin. 

\subsection{Exciting modes based on azimuthal number}

\begin{figure*}
\includegraphics[width=1\linewidth]{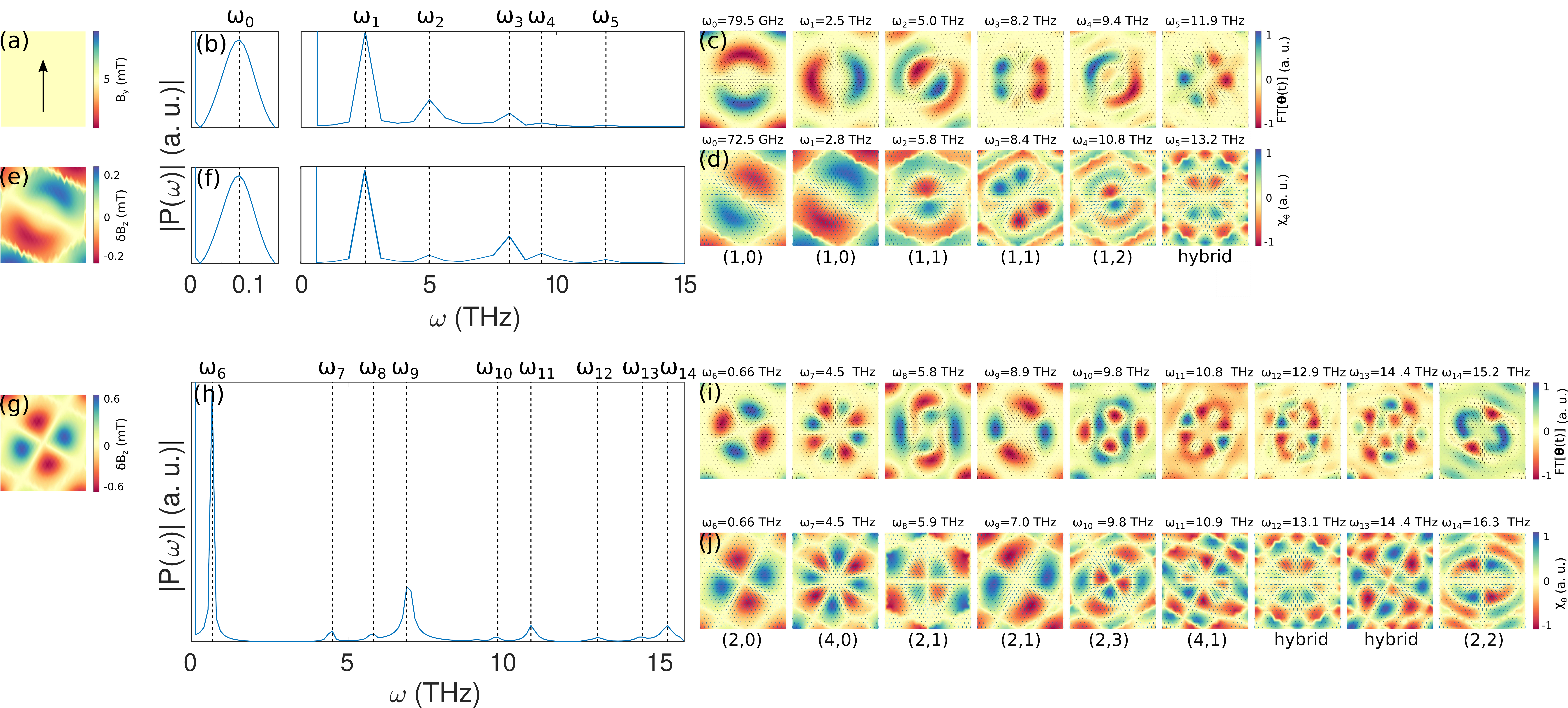}
\caption{Dynamical response of the skX of Pd/Fe/Ir(111) to a gaussian pulse in magnetic field.  (a, e, g) Spatial profile of the applied field along either $y$ (in plane) or $z$ (out of plane). (b, f, h) Corresponding Fourrier transform of the system's dynamics resolved up to 16~THz, where the position of peaks is indicated by vertical dash lines. (c, i) Fourrier $\theta$ profiles at the peaks. Since the in-plane uniform (a), and (1,0) out-of-plane (e) field profiles excite the same modes, only the response to the uniform field is shown in (c). (d, j) Corresponding $\theta$ components of the matching eigenvectors computed as in Sec.~\ref{sec:zoology}.  In all spatial plots, the view is limited to one unit cell. }
\label{fig:dyn}
\end{figure*}

In order to dynamically excite the modes, we examine the response of the system to a gaussian pulse in magnetic field of the form $\mathbf{B}(t,\mathbf{r})=B_{0}e^{\left(-t/\tau\right)^2}f(\mathbf{r}) \hat{\mathbf{u}}_B$, where $B_{0}=5$~mT, $\tau=40$~fs, $f(\mathbf{r})$ is a function determining the spatial dependence of the field, and $\hat{\mathbf{u}}_B$ a unit vector pointing either in plane or out of plane. The results are gathered in Fig.~\ref{fig:dyn}. 

We first apply a uniform in-plane field with $f(\mathbf{r})=\mathrm{cst}$, and $\hat{\mathbf{u}}_B=\hat{\mathbf{e}}_y$ [Fig.~\ref{fig:dyn}a]. The spectral response of the system is shown in Fig.~\ref{fig:dyn}b, where the positions of the peaks are indicated by dashed lines, and arbitrarily labelled $\omega_{0-5}$.  
We identify a peak in the GHz range, and five more peaks in the THz range. 

The spatial distribution of spectral amplitudes in $\theta$ at each peak is shown in Fig.~\ref{fig:dyn}c.  We find slow and fast $(1,0)$ translation modes at respectively $\omega_0=$80~GHz and $\omega_1=$2.5~THz. The next peak at $\omega_2=5$~THz corresponds to the $(1,1)$ mode. Higher frequency modes are additional internal deformation modes with $l=1$, and a hybrid mode at around 12~THz. Based on Ref.~\onlinecite{mochizuki2012spin}, we can expect the modes at $\omega_{1,2}$ to be responsible for the gyration behavior when excited with an oscillating field. This will be investigated in Sec.~\ref{sec:gyration}.

In Fig.~\ref{fig:dyn}d, we match the spectral profiles in Fig.~\ref{fig:dyn}c with the corresponding $(l,p,\omega)$ eigenmodes computed as in Sec.~\ref{sec:zoology} with $\alpha=0.01$. We obtain a good agreement of the two methods, both in profiles and frequencies. This validates the use of the harmonic approximation for the lower frequency modes [Eq.~(\ref{eq:taylor_exp})]. 
We also carry out an additional simulation with an out-of-plane uniform field, and find a single resonnance peak at 1.88~THz corresponding to the $(0,0)$ breathing mode~\cite{mochizuki2012spin}, in good agreement with the harmonic approximation which predicts $\omega_{(0,0)}=2.07$~THz.

Next, we propose to show how a  magnetic field with a nonzero azimuthal number can excite the matching azimuthal modes of the skX.  In order to obtain an excitation profile that matches the periodicity of the skX state, the $\theta$ components of the eigenvectors are used as the spatial dependence of the field, i.e., $f(\mathbf{r})={\chi}_{\theta}$ and  $\hat{\mathbf{u}}_B=\hat{\mathbf{e}}_z$. 
We start with the $(1,0)$ field profile in Fig.~\ref{fig:dyn}e. The spectral response, given in Fig.~\ref{fig:dyn}f, shows that this yields a similar response to that of the uniform in-plane field, where the $l=1$ modes shown in Fig.~\ref{fig:dyn}c are once more excited.
Second, the $(2,0)$  profile in Fig.~\ref{fig:dyn}g is used, and yields the spectral response in Fig.~\ref{fig:dyn}h. We arbitrarily set the largest resolved  frequency  around 16~THz. In this interval, we identify 9 peaks, at frequencies which we label $\omega_{6-14}$. The corresponding spectral profiles and the matched up eigenvectors are respectively given in Figs.~\ref{fig:dyn}i and j.  We find that the majority of excited modes indeed pertain to the $l=2$ category. Additionally, some $l=4$ modes respond, as they also possess the 2-fold symmetry. This is especially true of the mode at $\omega_{11}$, which we previously classified as $(4,1)$. As touched upon in Sec.~\ref{sec:zoology}, it does not actually possess a 4-fold symmetry, and instead resembles a pair of superimposed $(2,1)$ modes.
 In this way, the $(l,p)$ classification reaches its limit at higher frequencies, where the number of nodes increases, and the pseudo-cylindrical symmetry found in lower frequency modes is broken due to the underlying hexagonal symmetry of the system. 

 \subsection{The gyration dynamics}\label{sec:gyration}

In skyrmion lattices,  two typically reported modes are the CCW and CW gyration modes~\cite{mochizuki2012spin,zhang2017eigenmodes,onose2012observation,okamura2013microwave,schwarze2015universal,ehlers2016skyrmion}. 
It has been shown that the center of a skyrmion can be viewed as a collective coordinate whose dynamics obeys Thiele's equation~\cite{kravchuk2018spin,makhfudz2012inertia}. 
In this case, the gyrotropic term should determine the sense of gyration based on the sign of the topological charge, and so the existence of both CCW and CW motion is not clearly understood. 

\begin{figure*}
 \includegraphics[width=.8\linewidth]{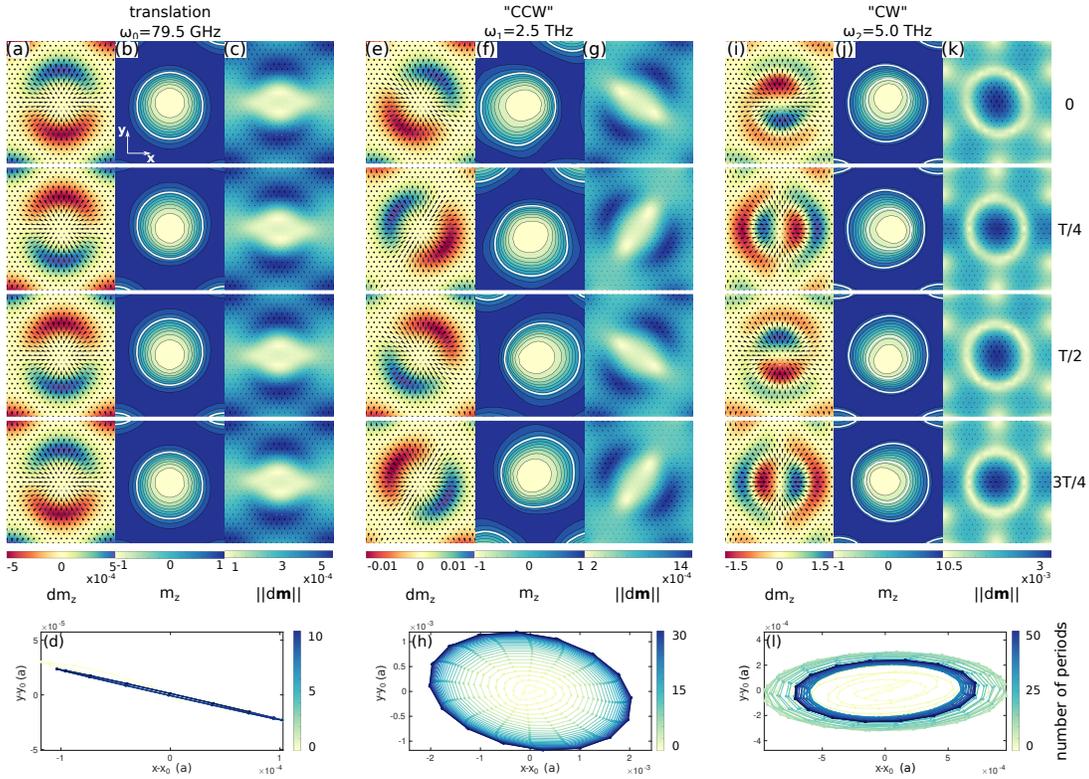}
\caption{Dynamical response of the skX of Pd/Fe/Ir(111) to an oscillatory in-plane field along $y$ at radial frequencies (a, b, c, d) $\omega_0$, (e, f, g, h) $\omega_1$, (i, j ,k, l) $\omega_2$. (a, e, i) Snapshots of the dynamics  over one period $T$ at an applied field amplitude of $5$~mT, where the black arrows represent the magnetization and the colorcode gives the deviation of the $m_z$ component from the ground state. (b, f, j) Snapshots of the dynamics at an applied field amplitude of $500$~mT  where the white isolines in $m_z$ correspond to the antinodes of the respective $dm_z$ profiles in (a, e, i). (c, g, k) Snapshots of the dynamics at an applied field amplitude of $5$~mT, where the  black arrows represent the deviation of the magnetization from the ground state, and the colorcode gives the amplitude of the deviation.  In every snapshot, the view is limited to one unit cell. (d, h, l) timetrace of the center of the skyrmion over a large number of periods at an applied field amplitude of $5$~mT, in which ($x_0,y_0$) correspond to the equilibrium position in units of the lattice constant.}
\label{fig:gyration}
\end{figure*}

Based on the system's response to the in-plane field [Fig.~\ref{fig:dyn}b], we focus on the modes previous labelled $\omega_{0,1,2}$, i.e., the two $(1,0)$ modes, and the $(1,1)$ mode. Following Ref.~\onlinecite{mochizuki2012spin}, we apply a spatially uniform oscillating in-plane magnetic field of the form $B_y(t)=B_{0} \cos(\omega_B t)$, with  $B_0=$~5mT or 500~mT.  The results are compiled in Fig.~\ref{fig:gyration}.

In Figs.~\ref{fig:gyration}a, e, and i, we show snapshots of the dynamics induced by a field amplitude of $5$~mT with $\omega_B$  respectively set to $\omega_0$, $\omega_1$, and $\omega_2$, over one period $T$. The black cones show the magnetic moments, and the colorcode gives the deviation of the $z$ component of the magnetization from the ground state. We find that at $\omega_0$, the amplitude of the deviation  remains mostly stationary while its sign oscillates [Fig.~\ref{fig:gyration}a], whereas at $\omega_{1,2}$, it respectively propagates CCW [Fig.~\ref{fig:gyration}e] and CW [Fig.~\ref{fig:gyration}i] over one period.  

For better visibility, we go beyond the linear response regime and increase the field amplitude to $500$~mT. We obtain the dynamics snapshots in Figs.~\ref{fig:gyration}b, f, and j, where we show contour plots corresponding to isolines in $m_z$, and the thicker white isolines match the position of the corresponding antinodes in Figs.~\ref{fig:gyration}(a, e, i). We find that the slow $(1,0)$ mode at $\omega_0$ induces apparent translation of the skyrmion [Fig.~\ref{fig:gyration}b], the fast $(1,0)$ mode at $\omega_1$ induces apparent CCW motion [Fig.~\ref{fig:gyration}f], and the $(1,1)$ mode at $\omega_2$ induces apparent CW antiphase motion of the skyrmion core and tail [Fig.~\ref{fig:gyration}j]. Note that the higher radial order of the CW mode was previously reported in Refs.~\onlinecite{zhang2017eigenmodes,mruczkiewicz2017spin}.

Next, in Figs.~\ref{fig:gyration}c, g, and k, we represent the deviation of the magnetization from the ground state configuration, $d\mathbf{M}(t)=\mathbf{M}(t)-\mathbf{M}_0$, as black cones,  where the colorcode gives the amplitude of $d\mathbf{M}(t)$. The field amplitude is reduced back to 5~mT. We find that in both $(1,0)$ modes, the deviation amplitude is essentially localized far from the skyrmion core, i.e., where $m_z>0$ [Figs.~\ref{fig:gyration}c and g]. On the other hand, most of the amplitude of the $(1,1)$ mode is localized onto the skyrmion core, with $m_z<0$ [Figs.~\ref{fig:gyration}k]. As made visible by the black cones, magnetic moments with $m_z>0$ $(<0)$ precess in the CCW (CW) direction, and this dictates the propagation direction of the perturbation.  We verified that this also applies to the modes at $\omega_{3-5}$ in Fig.~\ref{fig:dyn}c, where, at $\omega_3$,  more amplitude is found far from the core, and so the deviation propagtes CCW, while at $\omega_{4,5}$, most of the amplitude is localized onto the core, and the deviation propagates CW. Furthermore, at a high damping of $\alpha=1$, we can suppress the precession and recover a stationary perturbation amplitude with an oscillating sign, similar to the behavior described by Eq.~(\ref{eq:apply_mode}). As for the slow $(1,0)$ mode at $\omega_0$, it does not appear to be fundamentally different from the fast $(1,0)$ mode, but because the dynamics is much slower, it behaves like an overdamped mode where a small perturbation is damped down before it propagates.

Last, the time trace of the skyrmion center, defined as the center of mass of the topological charge distribution according to~\cite{kravchuk2018spin}, is shown in Figs.~\ref{fig:gyration}d, h, and l for 5~mT. The duration of the simulation is chosen as to allow the motion to reach an almost stationary state (blue lines). We find that the displacement of the skyrmion center over a period is consistently smaller than one atomic site. Within the linear regime, the skyrmion should thus be considered  stationary, and undergoing internal deformations. In this case, there is no gyrotropic term, as the center of mass has zero velocity in the atomistic framework. This conclusion remains true for an applied field of 500~mT.

In summary, we have found that the CCW and CW modes of the skX correspond respectively to the gapped $(1,0)$, and the $(1,1)$ modes. When these modes are excited by an oscillating magnetic field, the displacement of the center  of the skyrmion is negligible compared to the interatomic distance, and thus has zero velocity in the atomistic framework. The observed dynamics is therefore more akin to internal deformation than to gyration. The CCW or CW propagation direction of the perturbation was explained by the different spatial distribution of the mode amplitude, where the $(1,0)$ mode is localized far from the skyrmion core and thus the spins have $m_z>0$ and precess CCW,  while the $(1,1)$ mode is localized onto the skyrmion core, where the spins precess CW.


\section{Summary and perspectives}\label{sec:cl}


In this work, we computed the eigenmodes of skyrmion lattices in transition metal thin films. We  compared two classes of systems: systems where  noncollinear states are stabilized by the interplay of Heisenberg exchange and DMI, such as Pd/Fe/Ir(111), and systems in which noncollinear states are stabilized by exchange and the 4-spin interaction, such as Fe/Ir(111). 

First in Pd/Fe/Ir(111), we  found that the lowest frequency modes correspond to coupled internal deformation of the skyrmions. We suggested a classification based on azimuthal and radial numbers $(l,p)$, with $l\geq6$ and $p\geq 2$. The nature of the modes did not change with the inclusion of frustrated exchange and high-order terms, but the eigenfrequencies of modes with higher $l$ and $p$ increased slightly compared to the case with only effective exchange.

Second, in systems like Fe/Ir(111), we showed that the 4-spin interaction can stabilize a lattice of topological objets which are not skyrmions. In this case, the charge carriers do not exhibit internal degrees of freedom of the $(l,p)$ kind, and they are not separated by energy barriers. 
This demonstrates that the energy barriers that separate individual skyrmions do not automatically derive from the nontrivial topology, and neither do the internal degrees of freedom.
We note that in Ref.~\onlinecite{paul2020role}, the authors show that isolated skyrmions can be stabilized in Pd/Fe/Ir(111) and other similar systems at zero DMI, by the 4-spin interaction. However, these skyrmions exist as metastable excitations of the FM ground state, and in the absence of DMI, these systems do not exhibit noncollinear magnetic ground states--whether spin spirals or skX. That is because the 4-spin interaction has the opposite sign to that of Fe/Ir(111), and thus yields an energy maximum in the dispersion of $2Q$ spin spirals [Fig.~\ref{fig:dispersions}b].

Last, we performed magnetization dynamics simulations in the skX of Pd/Fe/Ir(111). We showed how the skX modes with a given $l$ can be selectively excited by a magnetic field with matching azimuthal number. We identified the CCW and CW modes as the gapped $(1,0)$, and $(1,1)$ modes. We showed that the dynamics resulting from their excitation under an oscillating magnetic field is an internal deformation propagating either CW or CCW, depending on whether the mode amplitude is localized onto the skyrmion core, or far from it.

We have shown that a nonuniform magnetic field  could selectively excite $l=1$ and $l=2$ modes based on their azimuthal number. Experimentally, a magnetic field carrying orbital angular momentum can be  generated by a Laguerre-Gauss electromagnetic beam~\cite{padgett2004light}. In our simulations, the field profile matched the periodicity of the underlying skX, which seems challenging to realize in practise. Nevertheless, the same principle could be applied to  selectively excite the modes of an isolated skyrmion. 
Alternatively, in  materials exhibiting both magnetic and ferroelectric orders, as is the case of Cu$_2$OSeO$_3$, the $l=2$ mode  is associated to an oscillating electric dipole moment, and could therefore be electrically excited~\cite{schutte2014magnon}. One can speculate that the other modes would exhibit a similar behavior. 
 
 So far, internal modes with $l \ge 4$ and $p \ge 1$ have rarely been reported for isolated skyrmions~\cite{lin2014internal}. Ref.~\onlinecite{zhang2017eigenmodes} is a good demonstration of how the $(1,1)$ mode, present in the skX phase, is absent in the isolated skyrmion, and it was speculated that the presence of this particular mode depends on interskyrmion interactions. However, it is possible that other types of confining potentials would have the same effect, as the $(1,1)$ mode was also reported in a skyrmion confined in a nanodot~\cite{mruczkiewicz2017spin}. An isolated skyrmion possessing these additional stable degrees of freedom would benefit from a large entropic stabilization effect~\cite{desplat2018thermal,von2019skyrmion,desplat2020path}, and would thus be very interesting for spintronics applications requiring a large thermal stability, such as data storing and processing~\cite{sampaio2013nucleation,fert2013skyrmions}.


\appendix


\section{Energy dispersions}\label{app:dispersions}

\paragraph{$1Q$ spin spirals} For single N\'eel-type spin spirals with wave vector $\mathbf{q}$,  the magnetization at lattice site $\mathbf{R}_i$ is given by,
\begin{equation}
\mathbf{m}^i=\mathbf{R}_q \cos\left( \mathbf{q} \cdot \mathbf{R}_i \right) + \mathbf{I}_q \sin\left( \mathbf{q} \cdot \mathbf{R}_i \right),
\end{equation}
where $\mathbf{R}_q=(0,0,1)$ and $\mathbf{I}_q=\mathbf{a}_1+\mathbf{a}_2$, with $\mathbf{a}_{1,2}=(\mp 1/2,\sqrt{3}/2,0)$,  the basis vectors for the monoatomic hexagonal unit cell.
 
\paragraph{$2Q$ spin spirals} Based on~\cite{heinze2011spontaneous}, we plot the energy of 2$Q$ spin spirals with wave vectors $\mathbf{q}_1$ and $\mathbf{q}_2$, where the magnetization at lattice site $\mathbf{R}_i$ is given by,
\begin{eqnarray}
 m_x^i & = & \cos\left( \mathbf{q}_2 \cdot \mathbf{R}_i \right) \sin\left(\mathbf{q}_1 \cdot \mathbf{R}_i\right), \\
m_y^i & = & \sin\left(\mathbf{q}_2 \cdot \mathbf{R}_i\right), \\
m_z^i & = & \cos\left( \mathbf{q}_2 \cdot \mathbf{R}_i \right) \cos\left(\mathbf{q}_1 \cdot \mathbf{R}_i\right).
\end{eqnarray}
 To make the $2Q$ states commensurate with the supercell, the unit cell must contain two Fe atoms, with base vectors $\mathbf{a}_1=(1,0,0)$ and $\mathbf{a}_2=(0,\sqrt{3}, 0)$ in direct space, and $\mathbf{b}_1=(1,0,0)$ and $\mathbf{b}_2=(0,1/\sqrt{3}, 0)$ in reciprocal space. For $\mathbf{q}_1 \parallel \mathbf{b}_1 \parallel \overline{\mathrm{\Gamma K}}$ and   $\mathbf{q}_2 \parallel \mathbf{b}_2 \parallel \overline{\mathrm{\Gamma M}}$, the state at $q_1=q_2=0.2$ is the multi-$Q_M$ star that ressembles the nanoskX state~\cite{heinze2011spontaneous}. In this configuration, the boundary of the first BZ for the biatomic unit cell is simultaneously reached along $ \overline{\mathrm{\Gamma K}}$ and $ \overline{\mathrm{\Gamma M}}$ for $q_1=q_2=0.5$.

\section{Deriving the eigenmodes of the dynamics}\label{app:extract_modes}

To obtain the eigenmodes of the dynamics, a harmonic expansion of the Hamiltonian in Eq. (\ref{eq:hamil}) is performed about the ground state configuration, $\boldsymbol{M}_0$, as,
\begin{equation}\label{eq:taylor_exp}
\mathcal{H}(\boldsymbol{M}) \approx \mathcal{H}_0 (\boldsymbol{M}_0) + \frac{1}{2} \big( \boldsymbol{M}-\boldsymbol{M}_0\big)^T H_{ \boldsymbol{M}_0}
\big( \boldsymbol{M}-\boldsymbol{M}_0\big),
\end{equation}
where $H_{ \boldsymbol{M}_0}$ is the Hessian matrix of the energy evaluated at  $\boldsymbol{M}_0$.

To solve the dynamics of small excitations about the ground state, we linearize the  Landau-Lifshitz-Gilbert (LLG) equation,
\begin{equation}\label{eq:LLG}
\dot{\mathbf{M}}= -\frac{1}{(1+\alpha^2) \hbar}  \left[  \mathbf{M} \times \frac{\partial \mathcal{H}}{\partial \mathbf{M} } + \alpha \left( \mathbf{M}\times \frac{\partial \mathcal{H}}{\partial \mathbf{M} } \right) \times \mathbf{M} \right],
\end{equation}
in which $\alpha$ is the dimensionless Gilbert damping, $\hbar$ is the reduced Planck constant, and a dot denotes a time derivative. This is done by injecting  Eq.~(\ref{eq:taylor_exp}) into (\ref{eq:LLG}). We choose polar and azimuthal angles $\theta$ and $\phi$ to describe the 2 degrees of freedom at each magnetic moment. The time evolution of small deviations from the ground state $\boldsymbol{\Theta} = \boldsymbol{\theta} - \boldsymbol{\theta}_0 $ and $\boldsymbol{\Phi} = \boldsymbol{\phi} - \boldsymbol{\phi}_0 $ then takes the form,
\begin{equation}\label{eq:linear_llg}
\begin{pmatrix}
{\boldsymbol{\dot{\Theta}}} \\
{\boldsymbol{\dot{\Phi}}} \\
\end{pmatrix}
=
\mathcal{T}_{ \boldsymbol{M}_0}
\begin{pmatrix}
\boldsymbol{\Theta} \\
\boldsymbol{\Phi} \\
\end{pmatrix},
\end{equation}
in which $\mathcal{T}_{ \boldsymbol{M}_0}$ is the transfer matric of the dynamics evaluated at ${\boldsymbol{M}_0}$.  More details on the derivation are given in~\onlinecite{desplat2018thermal}. 

 Next, the transfer matrix is diagonalized by solving the eigenvalue problem,
 \begin{equation}\label{eq:evalue_eq} 
 \mathcal{T}_{ \boldsymbol{M}_0} \boldsymbol{\chi} = \lambda \boldsymbol{\chi}.
  \end{equation}
  The $2N$ obtained eigenvalues are complex conjuguates of the form $\lambda=\left(\sigma_k \pm i \omega_k \right)$, where $k=1 \hdots N$ is the mode index, $\sigma_k$, $\omega_k \in \mathbb{R}$, and $i^2=-1$. We arbitrarily select the $N$ solutions with positive imaginary part. 
For stable modes, as is the case of all the modes at an energy minimum, we have $\sigma_k < 0$, and $|\sigma_k^{-1}|$ is a characteristic timescale of the mode, while $\omega_k$ is its radial frequency. 

Last, we can apply the $k$th mode to the magnetic ground state $\mathbf{M}_0$ as,
\begin{equation}\label{eq:apply_mode}
\mathbf{M}(t)= \mathbf{M}_0 \odot \left(\mathbb{I}+ A_0 \operatorname{Re}\left(\boldsymbol{\chi}_k\right) \right) e^{\sigma_k t} e^{i \omega_k t},
\end{equation}
where $\mathbb{I}$ is the identity matrix, $A_0$ is an arbitrary amplitude, and the $\odot$ symbol denotes an element-wise vector multiplication. In the rest of this work, we simply denote  $\operatorname{Re}\left(\boldsymbol{\chi}_k\right)$ as $\boldsymbol{\chi}_k$ for readability.

\begin{acknowledgments}
We thank J.-V. Kim, 
V. P. Kravchuk, M. Garst and W. Wulfhekel for enlightening discussions, and G. P. M\"uller and M. Hoffmann for their help with Spirit. This research was supported by the University of Li{\`e}ge under Special Funds for Research, IPD-STEMA Programme. 
\end{acknowledgments}

%
\bibliography{/home/louise/Dropbox/writing/latex/skyrmionbib}
\end{document}